# Nanoscale elucidation of Na,K-ATPase isoforms in dendritic spines

Thomas Liebmann[1,3], Hans Blom[2,3], Anita Aperia[1,3], Hjalmar Brismar[1,2,3]

1. Department of Women's and Children's Health, Karolinska Institutet, Solna, Sweden

2. Department of Applied Physics, Royal Institute of Technology, Stockholm, Sweden

3. Science for Life Laboratory, Advanced Light Microscopy, Solna, Sweden

## Abstract

**Background**

The dimensions of neuronal synapses suggest that optical super-resolution imaging methods are necessary for thorough investigation of protein distributions and interactions. Nanoscopic evaluation of neuronal samples has presented practical hurdles, but advancing methods are making synaptic protein topology and quantification measurements feasible. This work explores the application of Photoactivated Localization Microscopy (PALM) pointilistic super-resolution imaging for investigation of the membrane bound sodium pump, the Na,K-ATPase, in matured neurons.

**Results**

Two isoforms of the sodium pump (ATP1a1 and ATP1a3) were studied in cultured neurons using the PALM-compatible fluorescent proteins PAGFP and mEos. Nanoscopic imaging reveals a compartmentalized distribution of sodium pumps in dendritic spines. Several nanoclusters of pumps are typically found in the spine head and relatively few in the spine neck. The density of sodium pumps was estimated from a quantification of detected single molecules to 600-800 pump copies/$\mu m^2$ in the spine heads.

**Conclusions**

We have utilized PALM for dissection of nanoscale localization in mature cultured neurons and demonstrated similar topology and quantification estimates with PAGFP and mEos. PALM topology assessments of the sodium pump appeared similar to previous STED studies, though quantification estimates varied notably, implying that sample labeling strategies and choice of nanoscopic imaging method are critical factors for correct quantification.

# Background

The discrepancy between protein sizes and the resolution of classical light microscopy has compelled the advancement of imaging methodology. Recent techniques utilizing varying optical principles have successfully demonstrated resolution surpassing the diffraction limit (Dyba et al., 2003, Betzig et al., 2006, Rust et al., 2006, Sharonov and Hochstrasser, 2006, Hell, 2007, Huang et al., 2010). These super-resolved imaging methods have become valuable assets for describing the molecular complexity and distribution of interacting proteins within micro and nanodomains of varying biological samples. Protein organization within excitatory neuronal synapses, exemplifies the need for super-resolved investigation to elucidate protein interactions that shape their complex properties (Okabe, 2007, Bito, 2010).

Though no method exists without limitation, there has been some recent success applying optical imaging methods for both nanoscopic localization and quantification in neurons. Stimulated emission depletion (STED) microscopy and stochastic optical reconstruction microscopy (STORM) have effectively facilitated resolved distributions in neurons (Willig et al., 2006, Dani et al., 2010, Blom et al., 2012). Methods more compatible with live-cell imaging such as universal point accumulation in nanoscale topography (uPAINT) have also been utilized to estimate protein domains in synapses (Nair et al., 2013). Quantification of proteins in neurons has been performed with STED and STORM, though these methods are generally restricted to use of immunolabeling techniques, which can result in incomplete labeling, sample modification and loss of measurement accuracy due to antibody sizes (Blom et al., 2011, Nair et al., 2013). An alternative technique utilizing exogenous expression of photoactivatable FPs is known as photoactivated localization microscopy (PALM). It is typically a slower imaging method, but offers high precision without the use of antibodies and provides opportunities for molecular quantification (Greenfield et al., 2009, Scarselli et al., 2013, Specht et al., 2013).

In this study we outline a feasible approach toward super-resolution microscopy in neurons. We describe a neuronal sample preparation for investigating protein densities in mature synaptic domains and apply multiple super-resolution imaging approaches. As a sample protein, we have selected the Na,K-ATPase sodium pump, the principal energy transformer in the brain which is the essential protein responsible for maintaining the resting membrane potential and a major controller of intracellular ion homeostasis (Kaplan, 2002, Morth, 2011). We have previously estimated subcellular distribution and abundance of one sodium pump isoform, ATP1a3, in neurons with STED. Here we apply additional super-resolved imaging to two different pump isoforms, ATP1a1 and ATP1a3, while discussing the practical advantages of the different techniques (Blom et al., 2011). Using structured illumination microscopy (SIM), we access the dendritic distribution of the pump with high volumetric resolution (Gustafsson et al., 2008). We then turn to PALM for estimates of pump abundance in synaptic signaling domains within spines. Considering the increasing options for PALM probes used in localization microscopy, we have selected two of the commonly used fluorescent protein (FP) variants, PAGFP and monomeric Eos (mEosFP), for comparing precision and protein density measurement in cultured neurons (Patterson and Lippincott-Schwartz, 2002, McKinney et al., 2009, Zhang et al., 2012).

## Methods

### Cell Culture

Hippocampal neurons were cultured according to our previously reported protocol, with the following deviations (Liebmann et al., 2012). Cells were seeded at a density of $2.6 \times 10^4/cm^2$. Media added for routine changing contained an additional 0.5 mM L-glutamine (Sigma). Cells were transfected after 17 days in culture with Lipofectamine 2000 (Invitrogen) and fixated or analyzed 6-7 days after transfection. Samples were fixed for 4 min at RT in paraformaldehyde (4% in PBS, warmed to 37°C). After washing 3 times in PBS, coverslips were mounted with Immu-mount (Thermo Fisher Scientific) and left to solidify overnight before imaging.

**Constructs and Transfection**

A mammalian mEos2 vector was made by subcloning mEos2 from pRSETa mEos2 (Addgene plasmid 20341) into the EGFP-C1 vector backbone. Mutagenesis was performed according to Zhang et al. to convert mEos2 to mEos3.2 (Zhang et al., 2012). ATP1a1 and ATP1a3 plasmids were generated by subcloning the pump isoform into EGFP-C1, PAGFP-C1 (a kind gift from Dr. Jennifer Lippincott-Schwartz), mEos2-C1 and mEos3.2-C1vectors. The sequence for mCherry was subcloned into the EGFP-N3 vector backbone (replacing EGFP) before inserting the PSD-95 sequence as applied previously (Liebmann et al., 2012). Rat sequences of Na,K-ATPase isoforms and PSD-95 were used for subcloning. The calcium-sensitive GCaMP5 plasmid was acquired from Dr. Loren L. Looger (Addgene plasmid 31788). DNA plasmids were transfected with Lipofectamine 2000 (Life Technologies) according to manufacturer recommendations.

**Microscopy**

Super-resolution and confocal imaging was performed on a Zeiss 780LSM Elyra PS.1 system. Confocal scanning and structured illumination microscopy were done with a 63× (1.4 NA, oil) objective while PALM imaging was done with a 100x (1.46 NA, oil) objective. PALM imaging of PAGFP and mEosPFs was accomplished with 405 nm activation/conversion and 488 nm (100 mW) or 561 nm (100 mW) wide-field laser excitation, respectively. Excitation was consistently maintained at 15% for PAGFP and 20% for mEosFPs. Activation/conversion intensities were kept at minimal levels to restrict stochastic activation events to low and distinct single molecule densities, and then gradually increased to maintain sufficient activation/conversion during imaging. Lasers were kept at oblique angles (near-TIRF) to maintain low axial excitation thickness and enhance signal contrast. PALM images consisted of $5 \times 10^3$ acquisition frames of 50 ms and 100 ms for PAGFP and mEosFPs, respectively.  Detector EMCCD-gain on the Andor iXon DU 897 (512x512 pixels) was set at 100.  PSD-95 images were acquired immediately before PALM imaging by wide-field laser excitation at 561 nm.  SIM imaging was performed with 488 nm excitation (3%), an EMCCD-gain of 30, a grating period of 28.0 µm, 3 rotations, and averaging of 2 frames.  Calcium spikes were detected at 3 Hz acquisition using an inverted Zeiss Axiovert 200 with a 40× (1.3 NA, oil) objective equipped with an Andor iXon+ 897 EMCCD camera and an X-Cite lamp with 10% illumination intensity and a standard GFP excitation-dichroic-emission filter set (Semrock).

**Analysis**

PALM analysis was performed with the Zeiss ELYRA P integrated system software (ZEN 2011 SP2 Black). Peak intensity to noise ratio for molecular detection was evaluated at 4.5 and 6.0 for PAGFP and mEosFPs, respectively. Peak mask size was 3x3 pixels (300x300 $nm^2$ in the image focal plane).  Grouping maximum on times and off gap times (5 frames) were 250 ms and 500 ms  for PAGFP and mEosFPs, respectively. The grouping pixel radius for re-occurring and persistent events was set to 150 nm. Rendering was done with a Gaussian display mode, 10 nm/pixel and 0.5xPSF expansion factor. Drift correction was done with model based autocorrelation of whole image section registrations.  SIM analysis was generated with integrated Zeiss ELYRA S system software (ZEN 2011 SP2 Black) with the following parameters:  manual processing settings with SR frequency weighting (1.5), maximum isotropy and auto noise filtering.

## Results and Discussion

**Establishing neuron development**

To validate our model system, we first examined the morphological and structural development of the cultured primary neurons used in this study.  We cultivated neurons to the point of expected synaptic maturity to ensure optimal morphology and protein organization reflective of functional neurons. Cultures were consistently transfected with the respective DNA plasmids at day 17 in vitro and allowed to express for an additional 6-7 days before microscopic evaluation. We first assessed the functional maturity of the cultured neurons by monitoring real-time calcium activity with the calcium sensor GCaMP5 in absence of an activity stimulus (Akerboom et al., 2012). In a sample recording, calcium levels spiked with an average frequency of 0.3 Hz (Figure 1a,b), demonstrating spontaneous inter-neuronal synaptic communication. We next assessed the morphological synaptic maturity by expression of GFP-ATP1a3 together with PSD-95, a scaffold protein that exhibits a highly specialized localization within the postsynaptic density of excitatory synapses. The synaptic maturity of the cultured neurons is evident in the presence of dendritic spines and the distinct spine localization of PSD-95 clusters (Figure 1c). Note the clear separation of spine heads from the dendrite shaft similar to spine development described in intact tissue and living animals.

Confocal images of the sodium pump confirm the dendritic expression in dendrite shafts and spines, but they also reveal the limited ability to resolve the relative differences of ATP1a3 abundance between the two regions, the distinct protein organization within the respective compartments, or the membrane association within these small structures. To examine the membrane localization of the sodium pump isoforms, we applied SIM with lateral and axial resolution of approximately 100 nm and 275 nm, respectively, verified with fluorescent nanobead calibration. With improved axial sectioning, we demonstrate the expected high level of membrane localization of both ATP1a1 and ATP1a3 (Figure 2a,b) (Azarias et al., 2013). Stack projections reveals an increased abundance in spine heads (Figure 2c) in contrast to a lower abundance in the thin spine neck.

To further improve the precision of the pump distribution in neurons, we selected the PALM pointilistic super-resolution approach. Though a majority of single molecule super-resolution studies suggest fresh sample preparation and TIRF imaging protocols, this is less ideal for dissociated neuronal cultures where a majority of the cell structure is far from the glass surface. Instead we have conveniently applied conventional fixation and mounting procedures to our cultured neurons, allowing sample preservation as done with confocal and SIM preparations. Sample fixation ensures that no subcellular movement of the investigated pumps occurs during the recording time. To maintain acceptable axial resolution and maximize signal contrast away from the glass surface, we excite the sample with oblique angle or near-TIRF illumination. With the described sample preparation and experimental setup, PALM provides a protein distribution in neurons with nearly the highest precision possible with light microscopy without the inherent localization accuracy loss from antibody labeling techniques. Our first PALM investigation utilized the photoactivatable variant of GFP (PAGFP) to demonstrate super-resolved imaging with a well-established, inert class of (FPs). PAGFP was fused to each of the pump subunits of interest and expressed in culture together with mCherry-tagged PSD-95. Overlay of wide-field images of PSD-95 and PALM images allows a comparison of super-resolved distributions between synaptic and dendrite regions. Aggregates of both isoforms were associated with synaptic PSD-95 (Figure 3a-c). PALM images show, in agreement with SIM, that spine necks often appear to contain lower levels of the sodium pump compared to dendrites and spine heads. This is also in agreement with results previously found with STED super-resolution imaging of endogenous sodium pumps (Blom et al., 2011).

The last years have seen development of an extensive palette of FPs with varying colors and on/off or color switching characteristics (Lippincott-Schwartz and Patterson, 2009, Stepanenko et al., 2011). Using a second, unrelated FP allows us to examine the influence of the probe on localization measurements. We selected green to red photoconvertible mEosFPs which offer the additional benefit of higher localization precision (McKinney et al., 2009, Zhang et al., 2012). Both mEos2 and mEos3.2 were tested and show no translocation or distribution differences compared to PAGFP (Figure 4a-b).

Inherent in pointilistic imaging is the summation of single molecule detection over hundreds or thousands of acquisition frames.  This facilitates quantification of molecule density within a given region of interest.  We have analyzed PALM images for both PAGFP and mEosFPs to compare molecule quantification with the two unrelated FPs.  Recordings were first analyzed for detection precision to validate the level of accuracy expected for each probe in our preparations.  Though data is often filtered for improved localization precision, our recordings have included all detection values.  Sample histograms show mean precision values of approximately 25 nm and 20 nm for PAGFP and mEosFPs, respectively (Figure 5a).  For quantification, spines were isolated from whole cell recordings and individually assessed for single molecules density calculations (Figure 5b).  Though important physical properties like extinction coefficient, quantum yield, photostability and activation/conversion efficiency are different for PAGFP and mEosFPs, our measured molecule densities are independent of used FP (Figure 5c) (McKinney et al., 2009).   This suggest that both probes can be utilized for quantitative molecular imaging at the nanoscale in neuron preparations with similar effectiveness.  The reported values should be considered as approximate as each of the probes has limitations preventing perfect detection.  Of primary interest is activation/conversion efficiency and reported photoblinking (Annibale et al., 2011).  In each case, some loss of accuracy could be expected for molecular counting and absolute density constructions. However, these studies may still provide novel biological insight (Lehmann et al., 2011).The distributions of Na,K-ATPase isoforms shown in Figures 3 and 4 were finally quantified on the single molecule level as shown in Figure 5b. The density for both isoforms was found to be 600-800 molecules/$\mu m^2$ in individual spines independent of used FP (Figure 5c).

An intrinsic advantage of using PALM to demonstrate super-resolved localization is the independence from antibodies, which can introduce localization inaccuracies from unspecific labeling and antibody displacement.  Additionally, neuronal labeling by transfection results in exogenous expression in some cells, which allows for single cell examination and eliminates contamination from presynaptic branches or neighboring cells. This is a concern when examining a protein like the Na,K-ATPase which exists in both presynaptic and postsynaptic membranes (Pietrini et al., 1992, Azarias et al., 2013). However, fusion proteins should always be used with the caveat that they provide exogenous protein and overexpression. With this in mind, we can consider our PALM deductions as overestimates of pump density in spines, as opposed to the previous STED measurements which are likely underestimates due to incomplete labeling (Blom et al., 2011). To perform absolute quantification of protein density, approaches where the endogenous expression of proteins is directly assessed has to be used. One such approach is to use transgenic animals with knock-in of a FP fusion to the protein of interest. This has recently been demonstrated with the scaffold protein Gephyrin (Specht et al., 2013). Interestingly it was found that endogenous and exogenous expressions are balanced, resulting in the same density of Gephyrin in dendritic clusters independent of amount of overexpression.

Despite the limitations in quantification that antibody labeling in STED or exogenous expression in PALM imaging gives, we can estimate the number of sodium pumps in a dendritic spine to be between less than one hundred to up to eight hundred. Is this a realistic number and is it sufficient for physiological sodium (and potassium) regulation in dendritic spines? For a potential change of 100 mV to occur in 20 ms across the membrane of a spine head with a 500 nm diameter ($\approx$1 $\mu m^2$ membrane area) and a membrane conductance of 0.01 F/$m^2$, approximately 250 sodium ions enter the spine. As the sodium pump may extrude three sodium ions per turnover, about 80 pumps are thus needed to restore intracellular sodium (assuming a restricted diffusion due to the spine neck) in one pump cycle. The estimated amount of sodium pumps thus seems to be more than enough for a single transient membrane potential spike. Trains of excitatory signaling may raise the sodium levels in dendritic spines by factors more than 10 times this amount (Rose, 2002). The number of sodium pumps previously estimated with STED might here be an underestimate for a fast regulation, while PALM likely provides an overestimate. The true number of sodium pumps is between those estimates, which is a sufficient number of pumps to give a capacity to rapidly restore sodium levels and prevent the postsynaptic spine compartment from being exposed to a prolonged high sodium concentration.

The second topological elucidation the nanoscopic data deliver is the low abundance of pumps in the neck region. We have previously seen the same arrangement using STED microscopy and speculate that this may have implications for the chemical isolation of signals between individual spines (Blom et al., 2011, Blom et al., 2012).

## Conclusions

Current super-resolution imaging methods provide a means of investigating protein organization at a level not previously feasible with optical techniques. In this study, we demonstrated the applicability of different FPs for dissecting the nanoscopic localization of proteins in 'typical' neuron preparations. Using PALM, we imaged the distribution of the Na,K-ATPase in cultured hippocampal neurons at 25 nm and 20 nm precision with PAGFP and mEosFPs, respectively. Our results show similar pump topology to that presented previously with STED, an alternative super-resolution technique which allows endogenous protein localization.

Though quantification with super-resolution techniques has seen limited use, we have shown efficacy of PALM for molecular counting estimates, a method which will see further advancement with improved photoactivatable FPs and more understanding of photoconversion and blinking characteristics. Density estimates results were similar between different PALM probes, but they produced considerably different results from antibody-dependent STED measurements on endogenous sodium pumps (<100 pumps), highlighting inherent differences in those nanoscopic approaches. With all of the discussed methods there are clear advantages and limitations, and selection of the correct approach is dependent on specific aims such as volumetric resolution and controlled expression levels as well as factors including the ability to fuse fluorescent labels to a protein of interest or availability of specific antibodies.

In the end, super-resolution imaging approaches allow investigation of protein distributions and even quantification that may resolve important biological questions or introduce new concepts of functional interaction in the brain. It remains to be shown how the sodium pump distribution alters resting membrane potential dynamics or how it can be modulated on the individual spine level, but this study contributes quantitative assessment of expression in spines that may add insight into the capacity of pump variants to regulate intracellular sodium.

**Competing Interests**

The authors declare that they have no competing interests.

**Authors' Contributions**

TL conceived the study, designed constructs, generated neuron cultures, performed imaging experiments, processed and analyzed images, generated figures and wrote the manuscript. HAB participated in design and coordination of the study, performed calibration measurements, contributed to analysis and interpretation and wrote the manuscript. HJB participated in design of the study, contributed to presentation of data and wrote the manuscript.

**Acknowledgements**

This project was funded by the Swedish research council (VR 2010-4270), Tornspiran Foundation and Märta och Gunnar V Philipsson foundation.

# Figure Legends

**Figure 1. Spontaneous activity and mature synaptic morphology.** **(a)** Cultured hippocampal neuron (24 DIV) expressing GCaMP5 (green). Scale bar = 25 µm. **(b)** Intensity recording of calcium-sensitive GCaMP5 from cell shown in (a), recorded at 3 Hz. **(c)** Confocal image of GFP-ATP1a3 (green) and PSD-95-mCherry (red) (23 DIV) showing mature cell morphology and developed spine structures. Subset images show magnified crop of the individual channels and the merged image. Scale bar = 25 µm.

**Figure 2. Nanoscopic resolution of Na,K-ATPase in spines with Structured Illumination Microscopy.** **(a)** Stack projection from SIM imaging of the sodium pump isoforms (ATP1a1 or ATP1a3) fused with GFP. Scale bar = 10 µm. **(b)** Single SIM slice from cells in (a) showing a cross-section of a dendrite branch.

Axial resolution measured to 275 nm. Scale bar = 2 µm. **(c)** Magnification of stack projection from (a). Scale bar = 2 µm.

**Figure 3. Nanoscopic resolution of Na,K-ATPase pumps in spines with PALM using PAGFP and PSD-95 synaptic labeling.** **(a)** Wide-field image of neuron dendrite expressing PSD-95 tagged with mCherry (red) to identify postsynaptic regions. **(b)** Constructed PALM image of PAGFP-tagged sodium pumps in rainbow color scale showing heterogeneous expression in dendrites and spines. The red line shows an outline of the dendrite branch constructed from a time projection of wide-field PAGFP images. **(c)** PALM image in gray scale overlaid with the corresponding PSD-95 image from (a). Scale bar = 500 nm.

**Figure 4. Nanoscopic resolution of Na,KATPase pumps in spines with PALM using mEosFPs.** Constructed PALM image of mEosFP-tagged sodium pumps in rainbow color scale showing heterogeneous expression in dendrites and spines. The red line shows an outline of the dendrite branch constructed from a time projection of wide-field mEosFP images. **(a)** PALM images of ATP1a1. **(b)** PALM images of ATP1a3. Scale bar = 500 nm.

**Figure 5. PALM measurement precision and quantification of Na,K-ATPase in spines.** **(a)** Histogram of unfiltered detection precision for single recordings of PAGFP-ATP1a1 or mEos3.2-ATP1a1 in hippocampal cells. Typical mean precision was estimated at 25 nm and 20 nm for PAGFP and mEos3.2, respectively. **(b)** Example of single spine PALM image with detected localization for mEos3.2-ATP1a3. Scale bar = 200 nm. **(c)** Boxplots of total molecular density in spines comparing PAGFP and mEosFPs for each of the Na,K-ATPase isoforms. Boxes present the interquartile range of each group and median value (red line). Significance test shows no difference of median values between different fluorescent proteins (two-sample t-test).

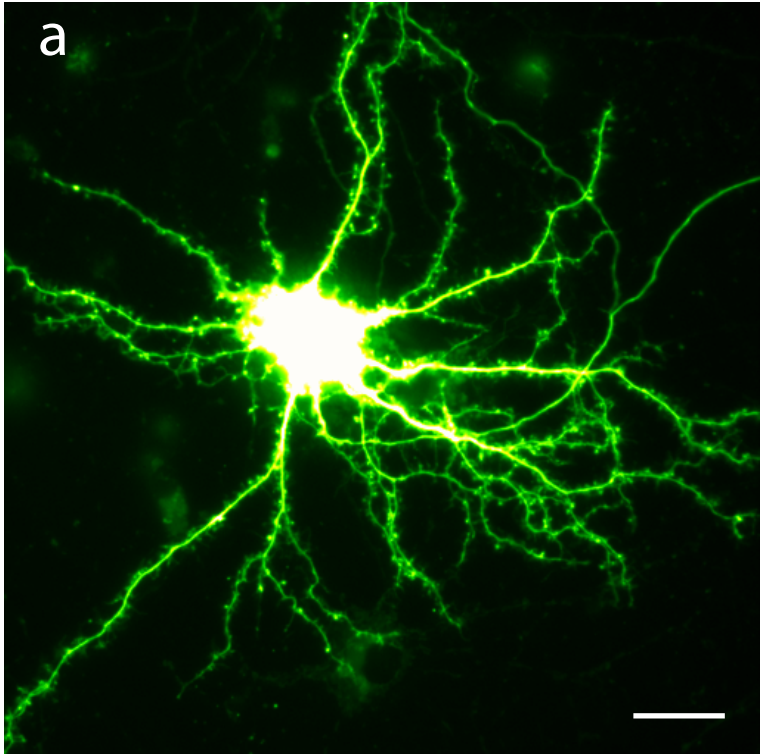
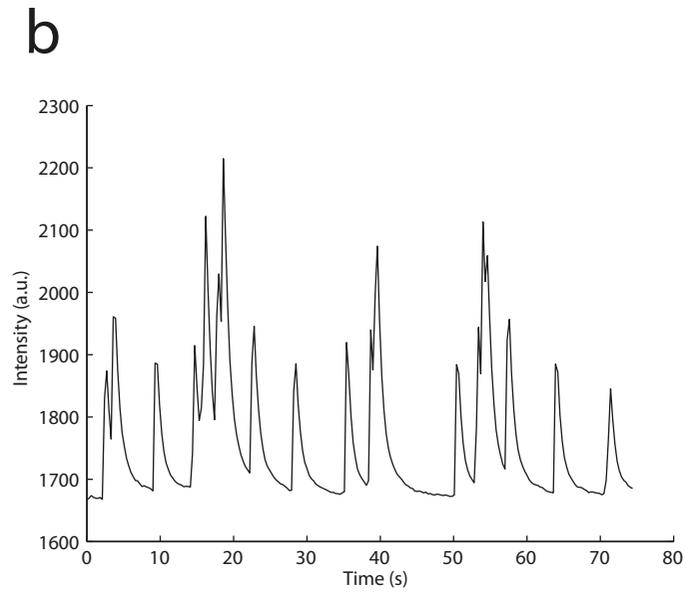
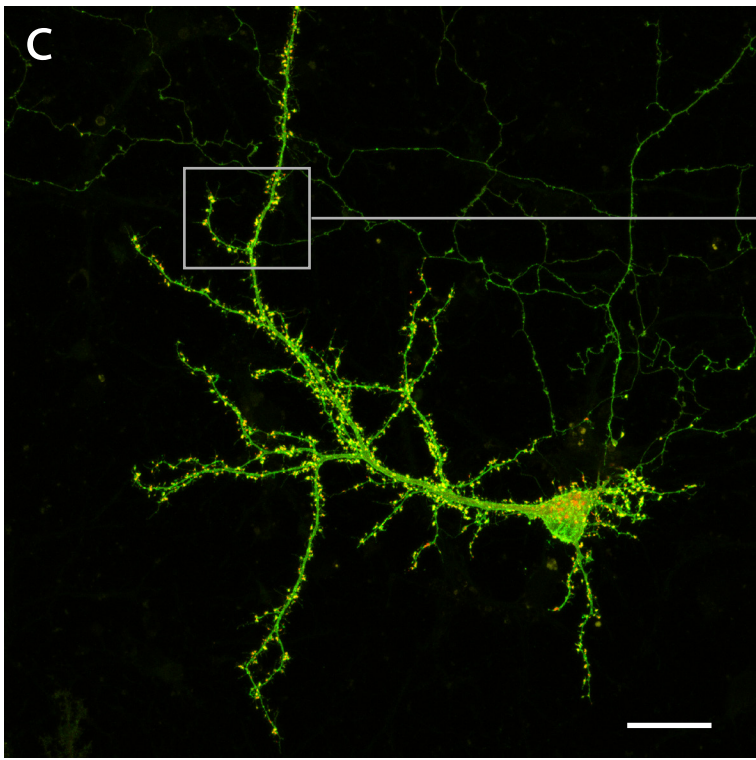
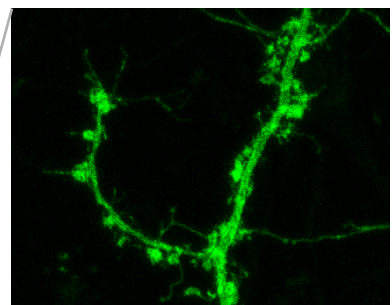
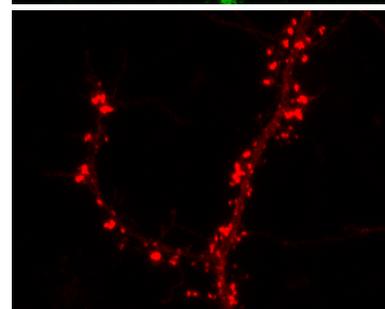
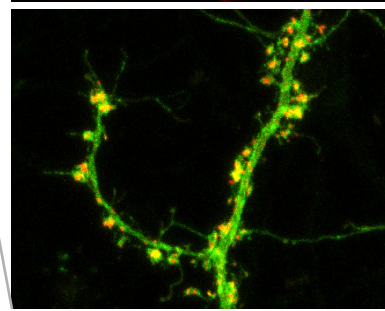

| | GFP-ATP1a1 | GFP-ATP1a3 |
|---|---|---|
| projection | 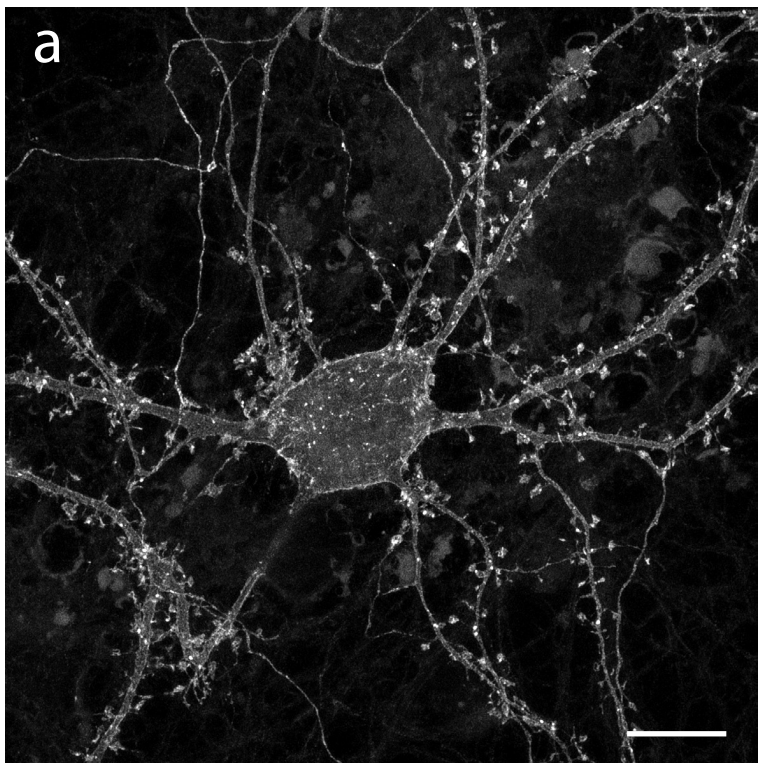 | 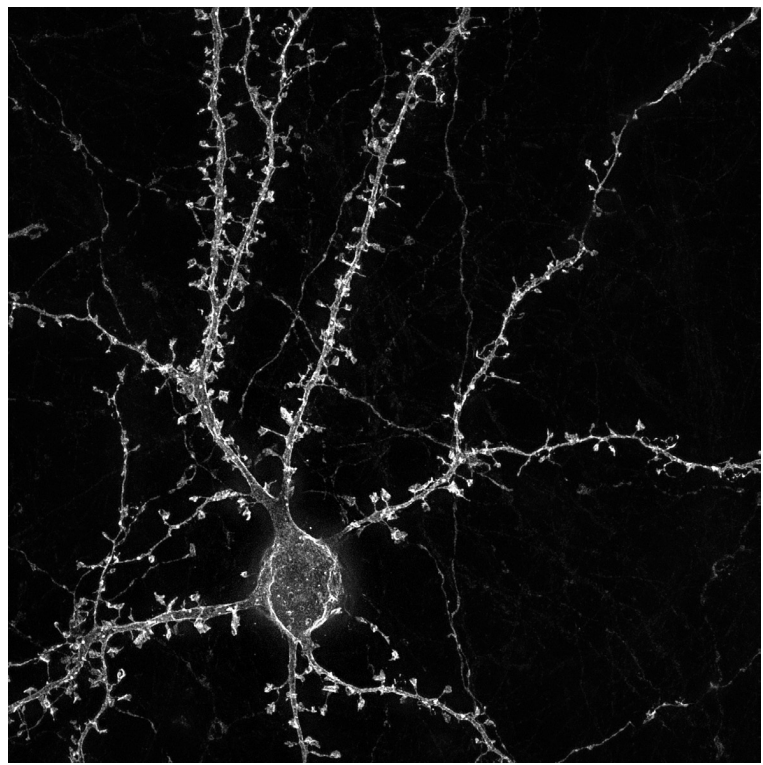 |
| optical slice | 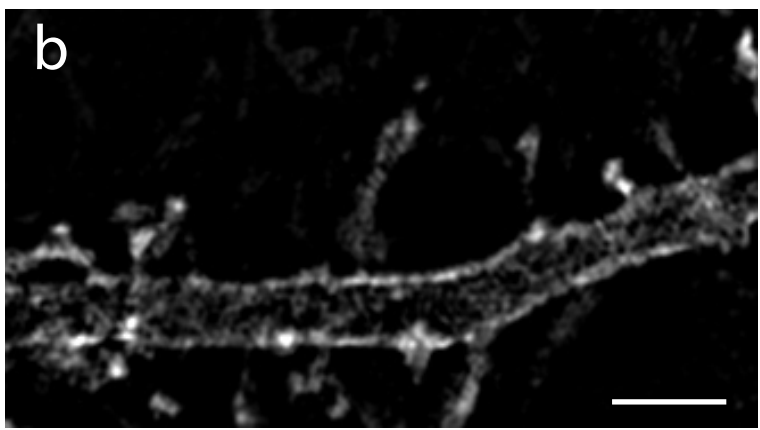 | 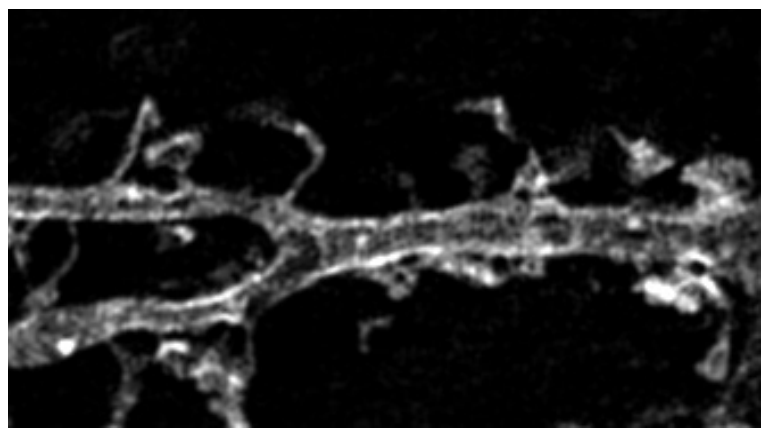 |
| projection | 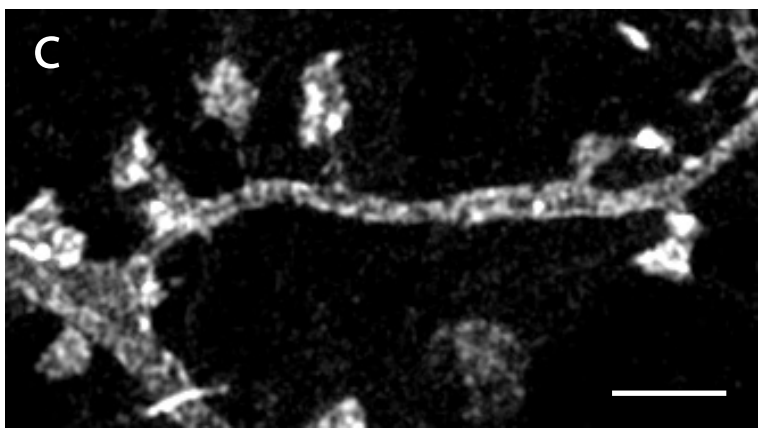 | 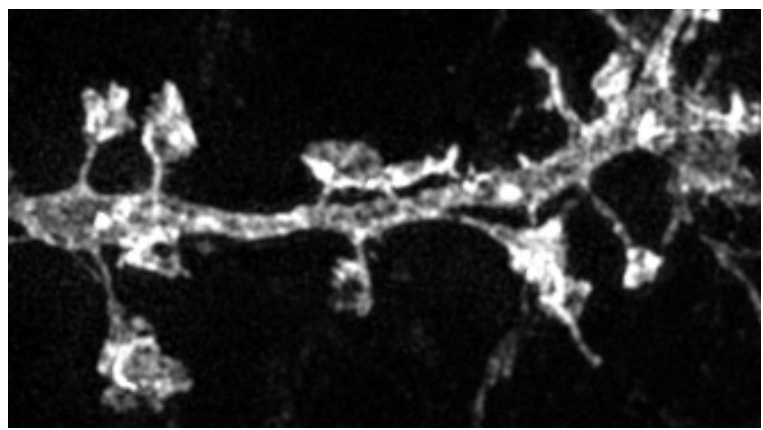 |

|  | ATP1a1 | ATP1a3 |
|---|---|---|
| PSD-95 (WF) | 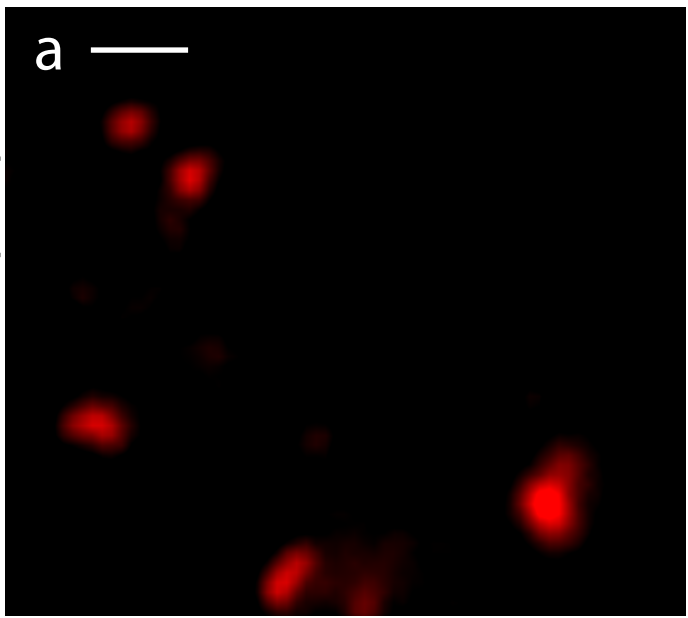 | 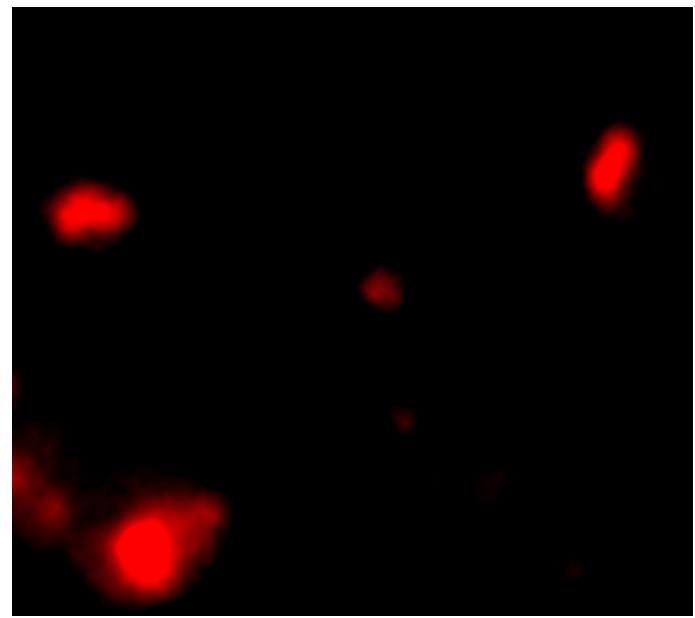 |
| PALM | 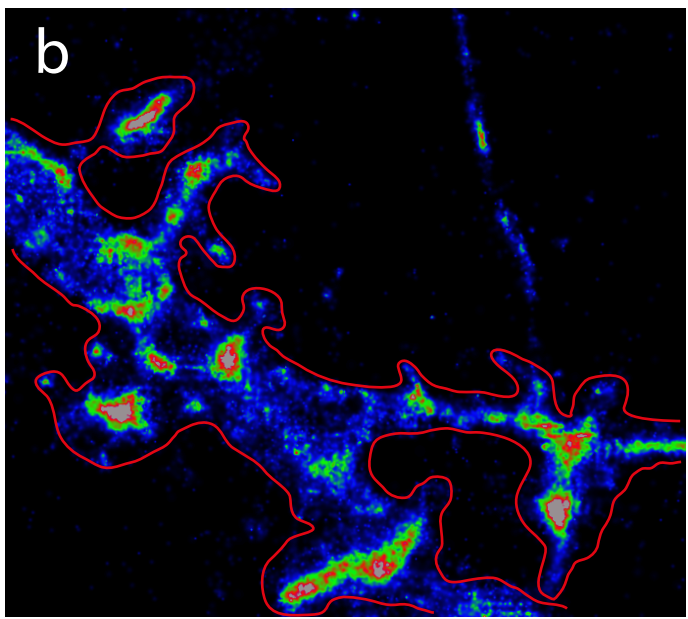 | 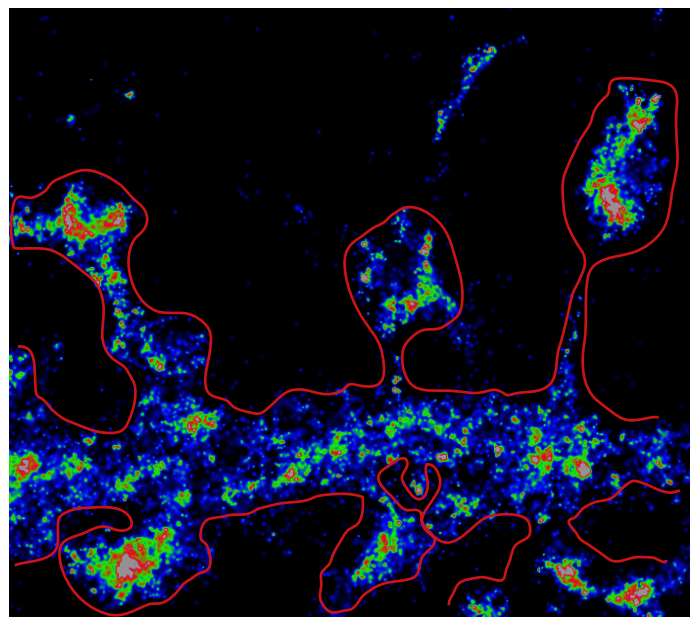 |
| PALM + PSD-95 | 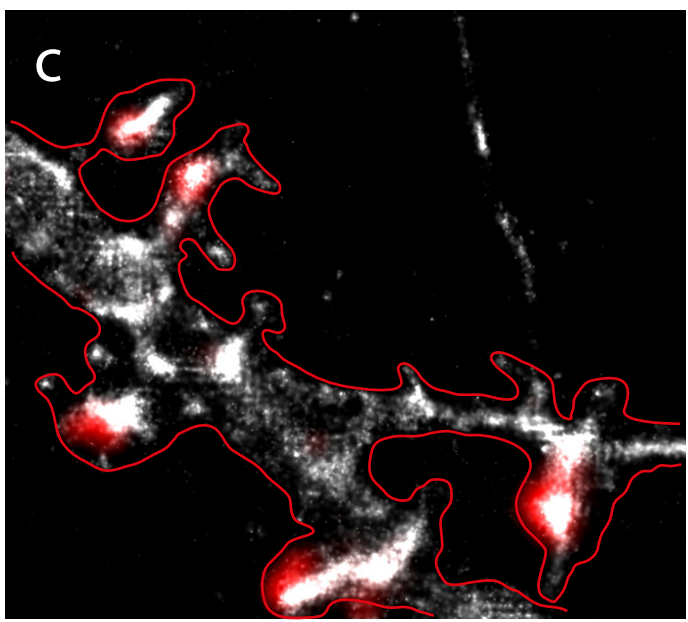 | 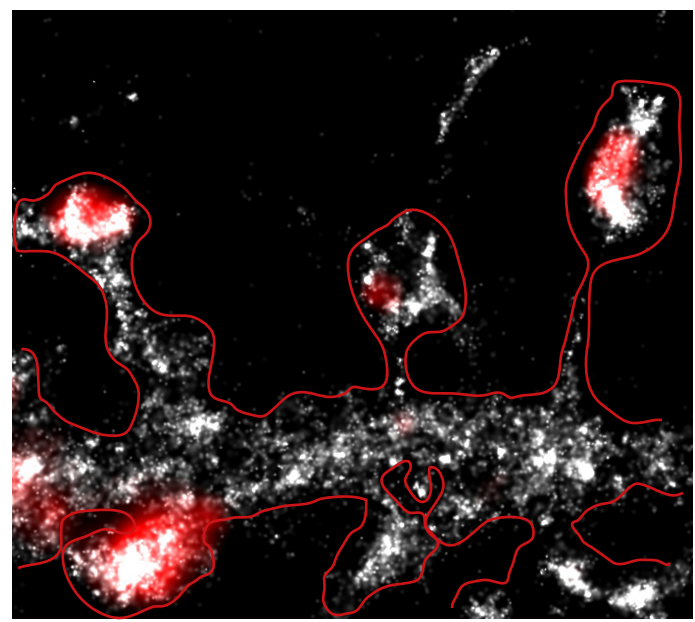 |

## ATP1a1 ## ATP1a3

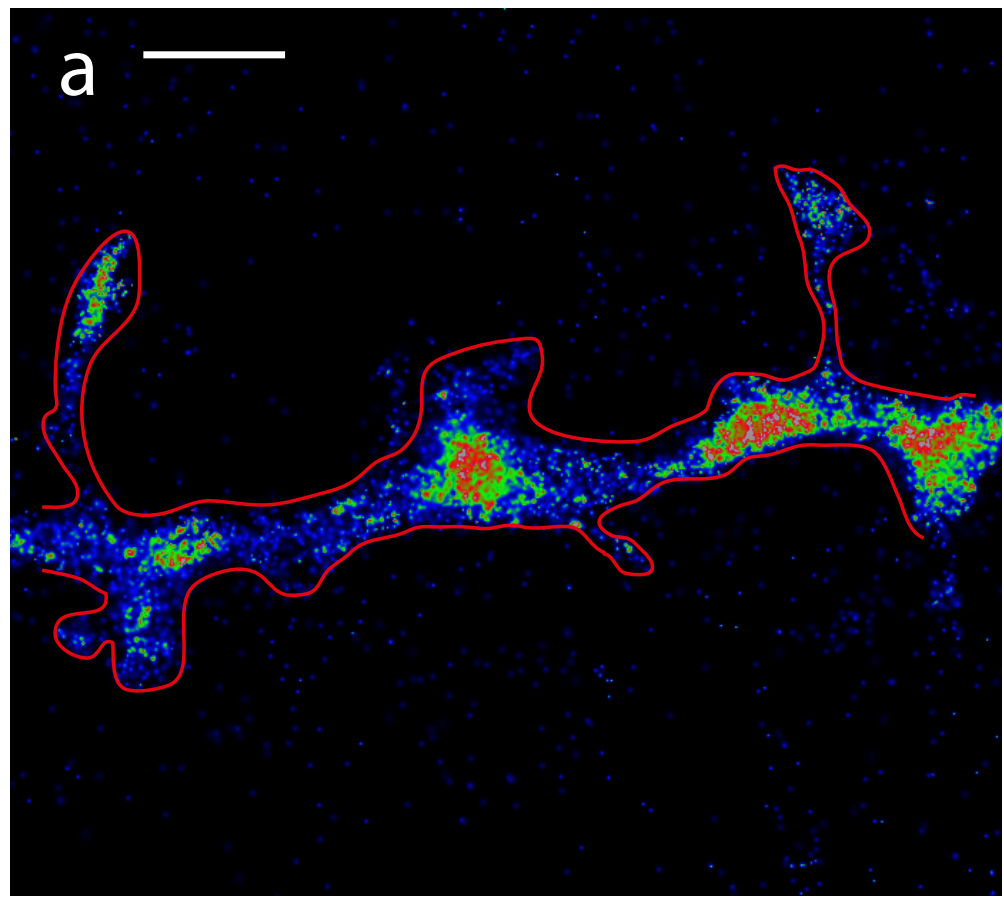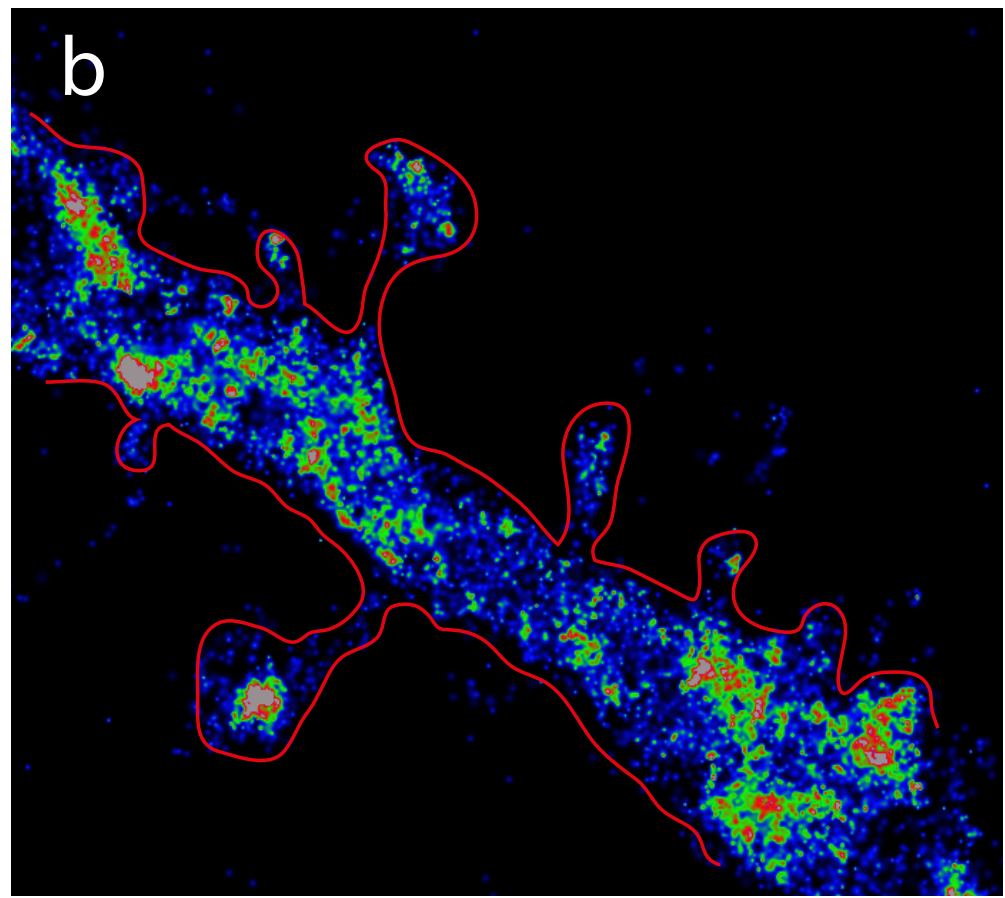

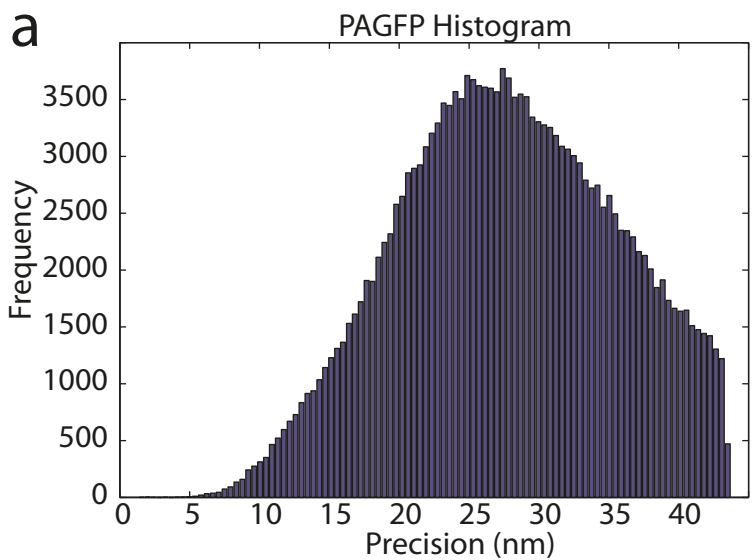
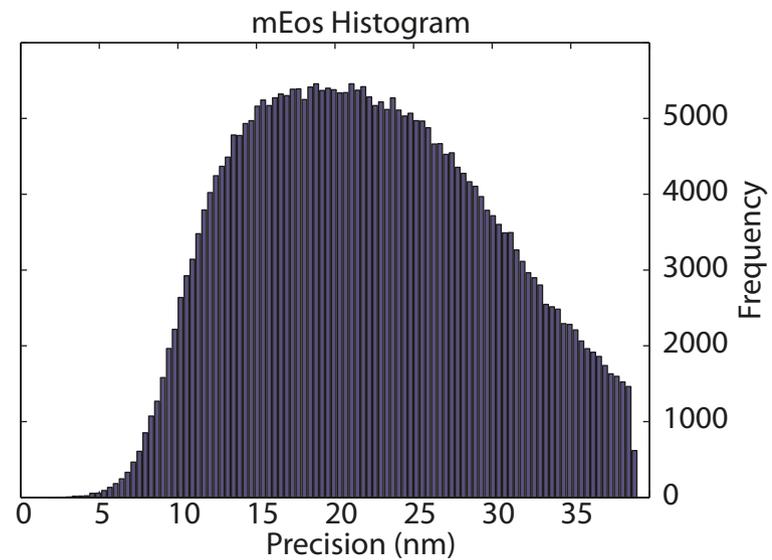
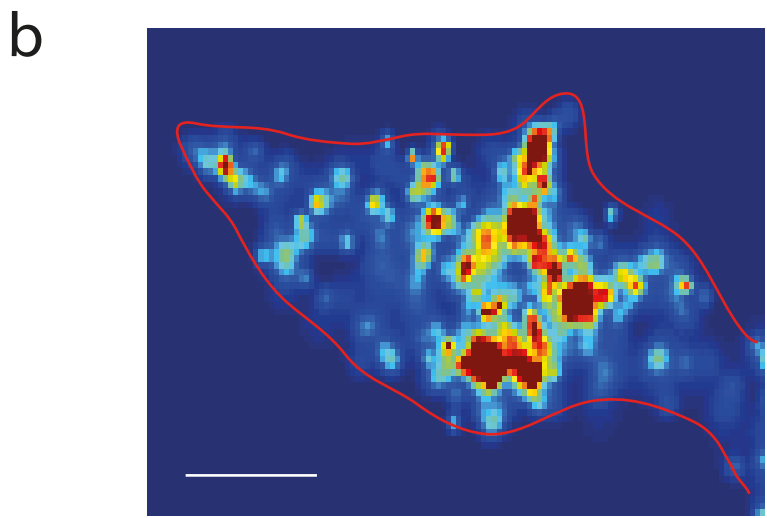
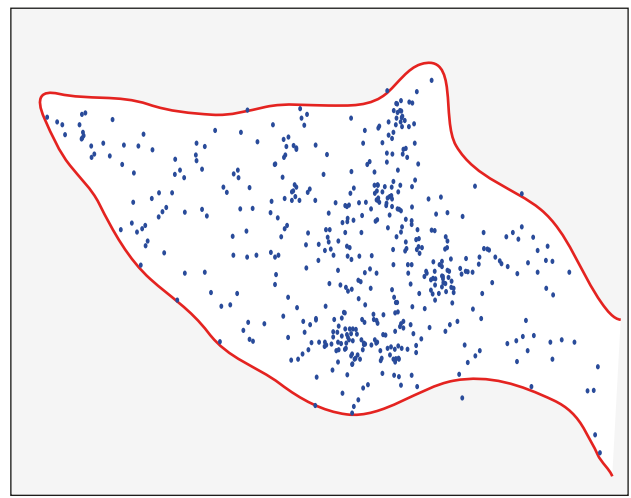
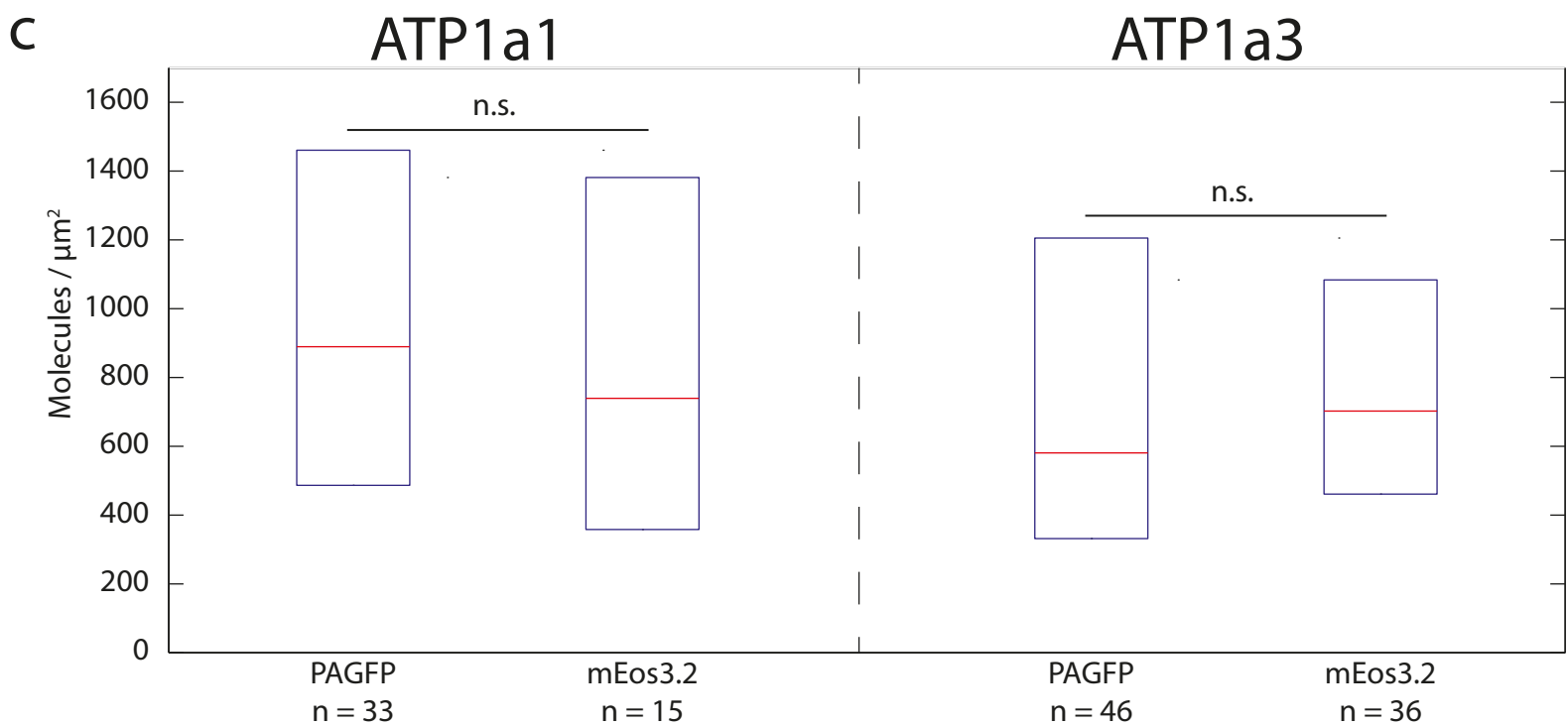